\documentclass[10pt,conference]{./IEEEtran}
\usepackage{amsmath,amssymb} 
\usepackage{graphicx}
\usepackage{bm}

\newtheorem{definition}{Definition}
\newtheorem{proposition}{Proposition} 

\newcommand{\qed}{\hfill $\square$ \par}

\newcommand{\rmd}{\mathrm{d}}
\newcommand{\rmi}{\mathrm{i}}

\newcommand{\llangle}{\langle\kern -.23em \langle}
\newcommand{\rrangle}{\rangle\kern -.23em \rangle}

\renewcommand{\vec}[1]{\boldsymbol{#1}}

\renewcommand{\baselinestretch}{0.90}

\begin{document}

\title{Generating Functional Analysis of Iterative Algorithms for Compressed Sensing}
\author{
  \authorblockN{Kazushi Mimura}
  \authorblockA{
    Department of Information Sciences, Hiroshima City University, \\
    Hiroshima 731-3194, Japan \\
    Email: {\tt\small mimura@hiroshima-cu.ac.jp}
  }
}
\maketitle
\begin{abstract}
 It has been shown that approximate message passing algorithm is effective in reconstruction problems for compressed sensing. 
To evaluate dynamics of such an algorithm, the state evolution (SE) has been proposed. 
If an algorithm can cancel the correlation between the present messages and their past values, 
SE can accurately tract its dynamics via a simple one-dimensional map. 
In this paper, we focus on dynamics of algorithms which cannot cancel the correlation and evaluate it by the generating functional analysis (GFA), 
which allows us to study the dynamics by an exact way in the large system limit. 
\end{abstract}

\section{Introduction}
\par
Dynamics of iterative algorithms for compressed sensing is discussed in this paper. 
We consider a problem that 
an $N$--dimensional vector $\vec{x}\in\mathbb{R}^N$ is reconstructed from 
an $M$--demensional ($M<N$) vector $\vec{y}\in\mathbb{R}^M$: 
\begin{equation}
  \vec{y} = A \vec{x}_0 + \vec{\omega}, 
\end{equation}
through an given $M \times N$ matrix $A\in\mathbb{R}^{M \times N}$. 
Here, $\vec{\omega} \in \mathbb{R}^M$ denotes a noise vector $\vec{\omega} \sim \mathcal{N}(\vec{0},\sigma_\omega^2 I)$. 
Since the ratio $\delta = M/N$, which is called the {\it compression rate}, is less than one, the system of equations undetermined. 
The original vector $\vec{x}_0$ may be however reconstructed if we have some knowledge of it, namely the sparsity. 
This problem \cite{Claerbout1973, Santosa1986, Donoho1989} is 
termed the reconstruction problem of compressed sensing \cite{Donoho2006,Candes2005,Candes2006,Candes2006b}. 
\par
To solve such undetermined systems, linear programming (LP) methods is widely applied and 
is investigated its performance \cite{Donoho2006, Candes2005, Candes2006, Candes2006b}. 
However, the LP might be still expensive to solve the large scale reconstruction problems. 
Recently, Donoho et al. have suggested an iterative algorithm which is called 
the approximate message passing algorithm (AMP) \cite{Donoho2009}. 
They have also proposed SE to evaluate its performance and 
have shown that the reconstruction performance of AMP is identical to that of the LP-based reconstruction \cite{Donoho2009}. 
Bayati and Montanari have provided the rigorous foundation to SE and 
have shown that SE can be applied to a general class of algorithms on dense graph, 
namely algorithms which can cancel the correlation between the present messages and its past values \cite{Bayati2010}. 
This correlation is often called a retarded self-interaction, which is caused by iterations, or the Onsager reaction. 
\par
Contrary to success of analysis for AMP, 
analysis for algorithms which cannot cancel the correlation between the present messages and their past values, 
e.g., the {\it iterative shrinkage-thresholding algorithm} (IST) \cite{Donoho2009, Zibulevsky2010}, is not discussed enough. 
In this case, we have to treat complex correlation. 
We focus on dynamics of algorithms which cannot cancel such correlation 
and evaluate the dynamics of IST by applying GFA \cite{DeDominicis1978, Heimel2001, Coolen2000, Coolen2005}. 
Dynamics, that appears in the information, has drawn attention so far 
\cite{Richadrson2001, Kabashima2003, Tanaka2005, Mimura2005, Mimura2006, Mimura2007, Mimura2009}. 
In GFA, we assume that the generating functional is concentrated around its average over the randomness in the large system limit, 
and we use the saddle-point methods to calculate the generating functional asymptotically. 
An advantage of GFA is to be able to evaluate dynamics of nonlinear systems exactly for the first few stages. 
To evaluate long time dynamics, approximation schemes may, on the other hand, have to be employed due to the computational cost. 
\par
This paper is organized as follows. 
The next section introduces reconstruction algorithms. 
Section III and IV explains about analysis and experiments, respectively. 
The final section is devoted to a summary.

\section{Settings and Algorithms}
\par
We assume the following to simplify the problem. 
Each element of the {\it original vector} $\vec{x}_0 =(x_{0,n}) \in \mathbb{R}^N$, is 
an i.i.d. random variable which obeys the distribution $p(x)=(1-\rho) \delta(x) + \rho (2\pi)^{-1/2} \exp(-x^2/2)$ 
with a given {\it signal density} $\rho \; (0 \le \rho \le 1)$, where $\delta(x)$ denotes Dirac's delta function. 
Each element of the {\it compression matrix} $A=(a_{mn}) \in \mathbb{R}^{M \times N}$ is 
an i.i.d. Gaussian random variable of mean zero and variance $M^{-1}$, i.e., $a_{mn} \sim \mathcal{N}(0,M^{-1})$. 
\par
Donoho et. al. have developped the following iterative algorithm achieving the performance of LP-based reconstruction. 
\begin{definition} \label{def:AMP}
  Starting from an initial guess $\vec{x}^{(0)}=\vec{0}$ and $\vec{z}^{(0)}=\vec{y}$, 
  the approximate message passing (AMP) algorithm iteratively proceeds by 
  \begin{eqnarray}
    \vec{x}^{(t+1)} &=& \eta_t(A^\top \vec{z}^{(t)} + \vec{x}^{(t)}), \\
    \vec{z}^{(t)} 
    &=& \vec{y} - A \vec{x}^{(t)} \nonumber \\
    & & + \frac 1{\delta} \vec{z}^{(t-1)} \langle \eta_{t-1}' (A^\top \vec{z}^{(t-1)} 
        + \vec{x}^{(t-1)}) \rangle. 
  \end{eqnarray}
  Here, $\{\eta_t\}$ is an appropriate sequence of threshold functions (applied componentwise), 
  $\vec{x}^{(t)} \in \mathbb{R}^N$ is the current estimate of the original vector $\vec{x}_0$, 
  $A^\top$ denotes the transpose of $A$ 
  and $\eta_t'(u)=\partial \eta_t(u) / \partial u$. 
  For a vector $\vec{v}=(v_1, \cdots, v_N)$, $\langle \vec{v} \rangle \triangleq N^{-1} \sum_{n=1}^N v_n$. 
  \qed
\end{definition}
One of other popular iterative algorithms \cite{Tropp2010} has the following form. 
\begin{definition} \label{def:IST}
  Starting from an initial guess $\vec{x}^{(0)}=\vec{0}$, 
  the iterative shrinkage-thresholding algorithm (IST) iteratively proceeds by 
  \begin{eqnarray}
    \vec{x}^{(t+1)} &=& \eta_t( \textstyle \frac 1c A^\top \vec{z}^{(t)} + \vec{x}^{(t)}), \label{eq:IST} \\
    \vec{z}^{(t)} 
    &=& \vec{y} - A \vec{x}^{(t)}. 
  \end{eqnarray}
  \qed
\end{definition}
Here, the parameter $c \ge 1$, which also appears in the separable surrogate functionals (SSF) method \cite{Daubechies2004}, 
is introduced to make IST be easy to converge. 
When $c=1$, the difference between AMP and IST is only whether the term 
$\frac 1{\delta} \vec{z}^{(t-1)} \langle \eta_{t-1}' (A^\top \vec{z}^{(t-1)} + \vec{x}^{(t-1)}) \rangle$ exists or not. 
The IST lacks this term which can cancel the correlation between the present messages and their past values. 
Due to this, the summation of massages cannot be regarded as a Gaussian random variable. 
It cannot therefore hope that the shrinkage-thresholding works properly. 
While the property of AMP is investigated theoretically and thoroughly \cite{Donoho2009, Bayati2010}, 
The dynamics of such algorithms for the reconstruction problem is not discussed enough so far.

\section{Analysis}
\par
The goal of our analysis is to evaluate the mean squared error (MSE) per component. 
\par
We analyze the dynamics in the large system limit where $N,M\to\infty$, while the compression rate $\delta$ is kept finite. 
The dynamics (\ref{eq:IST}) is a Markov chain, so the path probability $p[\vec{x}^{(0)},\cdots,\vec{x}^{(t)}]$, 
which is often called {\it path probability}, are simply given by products of the individual transition probabilities of the chain: 
\begin{align}
  p[\vec{x}^{(0)},\cdots,\vec{x}^{(t)}] 
  =& \delta [ \vec{x}^{(0)} ] \prod_{s=0}^{t-1} \delta [ \vec{x}^{(s+1)} \nonumber \\
   & - \eta_t ( \textstyle \frac 1c A^\top (\vec{y} - A \vec{x}^{(t)}) + \vec{x}^{(t)} + \vec{\theta}^{(t)}) ], 
  \label{eq:updating_rule}
\end{align}
which is called the {\it path probability}. 
Here, $\vec{\theta}^{(t)}$ is an external message which is introduced to evaluate the response function 
and these parameters $\{ \vec{\theta}^{0}, \cdots, \vec{\theta}^{(t)} \}$ are set to be zero in the end of analysis. 
The initial state probability becomes $p[\vec{x}^{(0)}]=\prod_{n=1}^N \delta[x_n^{(0)}]$. 
Therefore, we can calculate an expectation with respect 
to an arbitrary function ${\cal G}={\cal G}(\vec{x}^{(0)},\cdots,\vec{x}^{(t)})$ of tentative decisions as 
$\mathbb{E}_{\vec{x}}({\cal G}) \triangleq$ 
$\int_{\mathbb{R}^{(t+1)N}} ( \prod_{s=0}^t d\vec{x}^{(s)} )$ 
$p[\vec{x}^{(0)},\cdots,\vec{x}^{(t)}] {\cal G}$, 
where $\vec{x}$ denotes a set $\{\vec{x}^{(0)},\cdots,\vec{x}^{(t)}\}$ 
and $\mathbb{E}_X$ denotes the expectation with respect to a random variable $X$. 
To analyze the dynamics of the system we define the following functional that is called the {\it generating functional}. 
\begin{definition}
  The generating functional $Z[\vec{\psi}]$ is defined by 
  \begin{align}
    Z[\vec{\psi}] \triangleq 
    \mathbb{E}_{\vec{x}} \biggl( 
      \exp \biggl[ - \rmi \sum_{s=0}^t\vec{x}^{(s)}\cdot\vec{\psi}^{(s)} \biggr] 
    \biggr) ,
    \label{eq:def_Z}
  \end{align}
  where $\vec{\psi}^{(s)}=$ $(\psi_1^{(s)},$ $\cdots,$ $\psi_N^{(s)})^\top$. 
  \qed
\end{definition}
In familiar way \cite{DeDominicis1978, Coolen2000, Mimura2005}, 
one can obtain all averages of interest by differentiation, e.g., 
\begin{align}
  i\lim_{\vec{\psi}\to\vec{0}}\frac{\partial Z[\vec{\psi}]}{\partial \psi_n^{(s)}}
  &= \mathbb{E}_{\vec{x}} (x_n^{(s)}), \\ 
  -\lim_{\vec{\psi}\to\vec{0}}\frac{\partial Z[\vec{\psi}]}{\partial \psi_n^{(s)} \partial \psi_{n'}^{(s')}} 
  &= \mathbb{E}_{\vec{x}} (x_n^{(s)} x_{n'}^{(s')} ), \\ 
  i\lim_{\vec{\psi}\to\vec{0}}\frac{\partial Z[\vec{\psi}]}{\partial \psi_n^{(s)} \partial \theta_{n'}^{(s')}} 
  &= \frac{\partial \mathbb{E}_{\vec{x}} ( x_n^{(s)} ) }{\partial \theta_{n'}^{(s')}}.  
\end{align}
from $Z[\vec{\psi}]$. 
We assume that the generating functional is concentrated to its average 
over the random variables $\{A, \vec{x}_0, \vec{\omega} \}$ in the large system limit, 
namely the typical behavior of the system depends only on the statistical properties of the random variables. 
We therefore evaluate the averaged generating functional 
$\bar{Z}[\vec{\psi}] = \mathbb{E}_{\vec{x}, A, \vec{x}_0, \vec{\omega}} 
( \exp [ - \rmi \sum_{s=0}^t\vec{x}^{(s)}\cdot\vec{\psi}^{(s)} ] ) $, 
where $\overline{[\cdots]}$ denotes an expectation over $\{A, \vec{x}_0, \vec{\omega} \}$. 
Evaluating the averaged generating functional, one can obtain important parameters which describe the algorithm performance. 
Namely, we can evaluate the overlap, which is also called the direction cosine, 
between he original vector $\vec{x}_0$ and the current estimate $\vec{x}^{(s)}$ 
and the second moment of the current estimate. 
Since $||\vec{x}_0 - \vec{x}^{(t)}||_2^2 = ||\vec{x}_0||_2^2 - 2\vec{x}^{(t)} \cdot \vec{x}_0 + ||\vec{x}^{(t)}||_2^2$, 
we can evaluate MSE from the overlap and the second moment. 
Here, $\vec{x}^{(t)} \cdot \vec{x}_0$ denotes the inner product between $\vec{x}^{(t)}$ and $\vec{x}_0$. 
One finds the following proposition. \\
\par
\begin{proposition} \label{proposition:IST}
  For IST with an arbitrary sequence of threshold functions $\{\eta_s\}_{s=0}^t$, 
  MSE per component $\sigma_t^2$ of the current estimate $\vec{x}^{(t)}$ can be assessed as 
  \begin{align}
    \sigma_t^2 
    &\triangleq N^{-1} \mathbb{E}_{\vec{x}, A, \vec{x}_0, \vec{\omega}}(||\vec{x}_0 - \vec{x}^{(t)}||_2^2) \nonumber \\
    &= \rho-2m^{(t)}+C^{(t,t)}, 
    \label{eq:MSE}
  \end{align}
  in the large system limit, i.e., $N\to\infty$, where the parameters are given as follows. 
  \begin{align}
    m^{(s)}    &= \llangle x_{0} x^{(s)} \rrangle, \label{eq:sp_m} \\
    C^{(s,s')} &= \llangle x^{(s)} x^{(s')} \rrangle, \\
    G^{(s,s')} &= \llangle x^{(s)}(\vec{R}^{-1}\vec{v})^{(s')} \rrangle \mathbb{I}(s>s'), 
  \end{align}
  where $\mathbb{I}(\mathcal{P})$ denotes an indicator function which takes 1 if the proposition $\mathcal{P}$ is true, 0 otherwise. 
  Here, the average over the effective path measure $\llangle \cdots \rrangle$ is given by 
  \begin{align}
    \llangle g(\vec{x},\vec{v}) \rrangle 
    \triangleq&  \mathbb{E}_{x_0} \biggl( \int {\cal D}\vec{v} \int_{\mathbb{R}^{t+1}} \biggl( \prod_{s=0}^{t-1} \! dx^{(s)} \biggr) g(\vec{x},\vec{v}) \; \delta [x^{(0)}] \notag \\
    & \times \prod_{s=0}^{t-1} \! \delta [ x^{(s+1)} - \eta_s(x_0 \hat{k}^{(s)}+v^{(s)}+(\vec{\Gamma} \vec{x})^{(s)}) ] \biggr), \label{eq:Dv}
  \end{align}
  where 
  ${\cal D}\vec{v} =$ $|2\pi \vec{R}|^{-1/2}$ $d\vec{v}$ $\exp [ -\frac 12\vec{v}\cdot \vec{R}^{-1}\vec{v}]$, 
  $\vec{R} =$ $c^{-2} (\vec{1}+(c \delta)^{-1} \vec{G}^\top)^{-1}$ $\vec{D}$ $(\vec{1}+(c \delta)^{-1} \vec{G})^{-1}$, 
  $\vec{\Gamma} =$ $c^{-1}(c-1)\vec{1}$ $+ c^{-1} (\vec{1}+(c \delta)^{-1} \vec{G})^{-1} (c \delta)^{-1} \vec{G}$ and 
  $\hat{k}^{(s)} = c^{-1} |\vec{\Lambda}_{[s]}|$. 
  Each entry of $\vec{D}$ is $D^{(s,s')} \triangleq \sigma_\omega^2 + \delta^{-1} [ \rho -m^{(s)}-m^{(s')}+C^{(s,s')} ]$ and 
  each entry of $\vec{\Lambda}_{[s]}$ is $\Lambda_s^{(s',s'')} = \delta_{s,s'}+(1-\delta_{s,s'})(\delta_{s',s''} + (c \delta)^{-1} G^{(s'',s')})$. 
  The terms $(\vec{R}^{-1}\vec{v})^{(s)}$ and $(\vec{\Gamma} \vec{\sigma})^{(s)}$ denote 
  the $s^{\rm th}$ element of the vector $\vec{R}^{-1}\vec{v}$ and $\vec{\Gamma} \vec{\sigma}$, respectively. 
  \qed
\end{proposition}
\par
Outline of derivation is available in Appendix A. 
The parameters $m^{(s)}$ and $C^{(s,s')}$ are referred to as the overlap and the correlation function, respectively. 
Especially, $C^{(s,s)}$ gives the second moment of the $s^\mathrm{th}$ estimate. 
In GFA, we extract a one-dimensional iterative process which is statistically equivalent to the original $N$-dimensional iterative process. 
The effective path measure $\llangle \cdots \rrangle$ is an expectation operator with respect to such a one-dimensional process. 
Proposition \ref{proposition:IST} entirely describe the dynamics of the system. 
The term $(\vec{\Gamma} \vec{\sigma})^{(s)}$ in (\ref{eq:Dv}) is called the retarded self-interaction or the Onsager reaction term.

\section{Experiments}
\par
To validate the results obtained above, we performed numerical experiments in $N=2,000$ systems. 
For sparse signed original vectors, the sequence of the threshold functions \cite{Donoho2009} is chosen as 
$\eta_t(x;\hat{\lambda}_t)=(x-\hat{\lambda}_t)\mathbb{I}(x>\hat{\lambda}_t)+(x+\hat{\lambda}_t)\mathbb{I}(x<\hat{\lambda}_t)$ 
with $\hat{\lambda}_t=\lambda \sigma_t / c$, 
where $\lambda$ is a threshold control parameter and $\sigma_t$ is MSE per component of the current estimate. 
In practice, we cannot use a true MSE, since we do not know the original vector. 
We therefore have to use an alternate value instead of the true MSE. 
The MSE on zeros $\hat{\sigma}_t^2 = \mathbb{E}_{\vec{x}_0}[ \mathbb{I} (x_{0,n}=0) (x_n^t)^2 | x_{0,n}=0]$, 
which is referred to as MSEZ, is one of useful alternate values which are easy to estimate. 
We set $\hat{\sigma}_0^2 = \rho$ for an initial value. 
\par
\begin{figure}[t]
  \begin{center}
    \includegraphics[width=.7\linewidth,keepaspectratio]{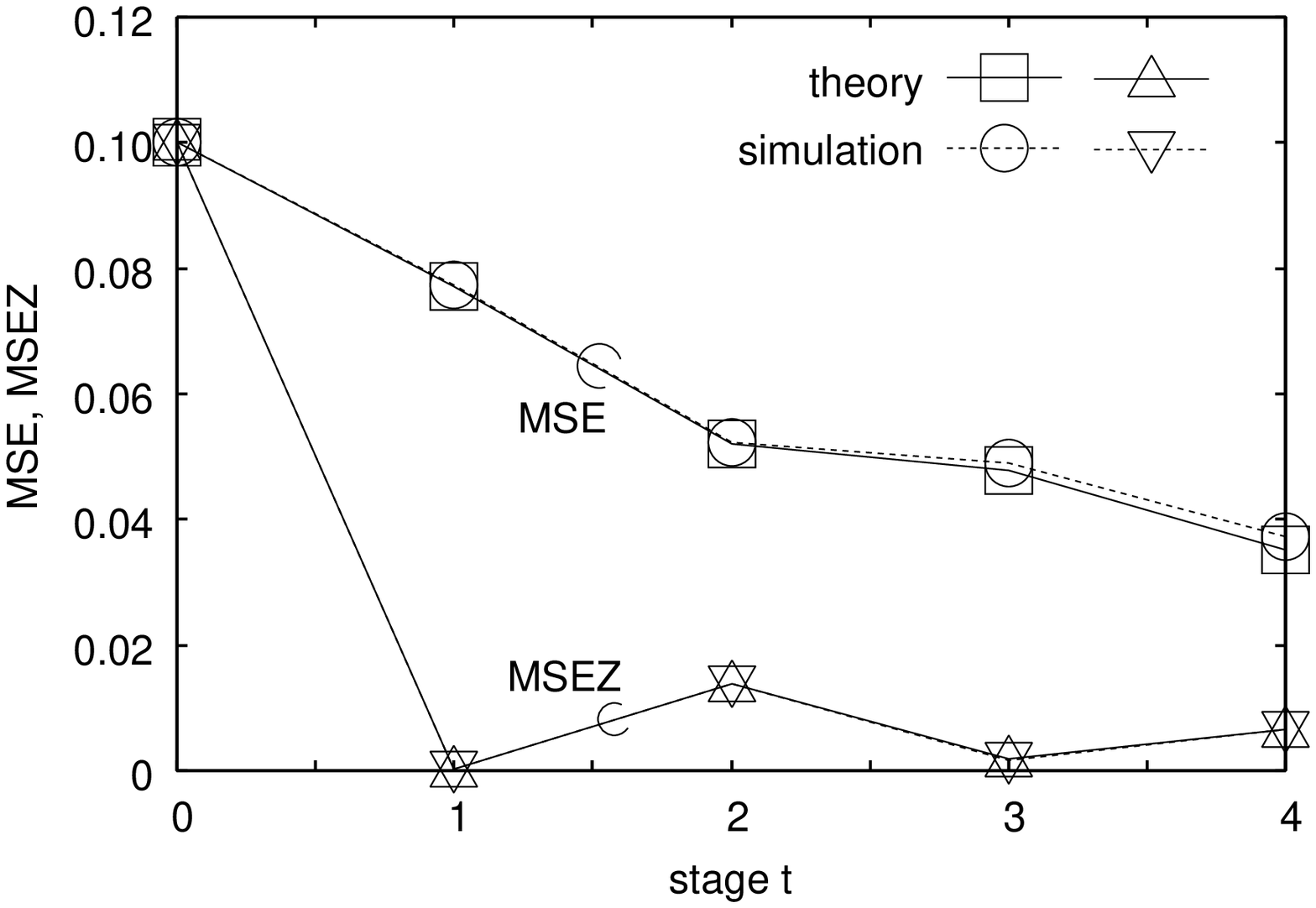} \\
    \footnotesize $\qquad$ (a) \normalsize \\[2mm]
    \includegraphics[width=.7\linewidth,keepaspectratio]{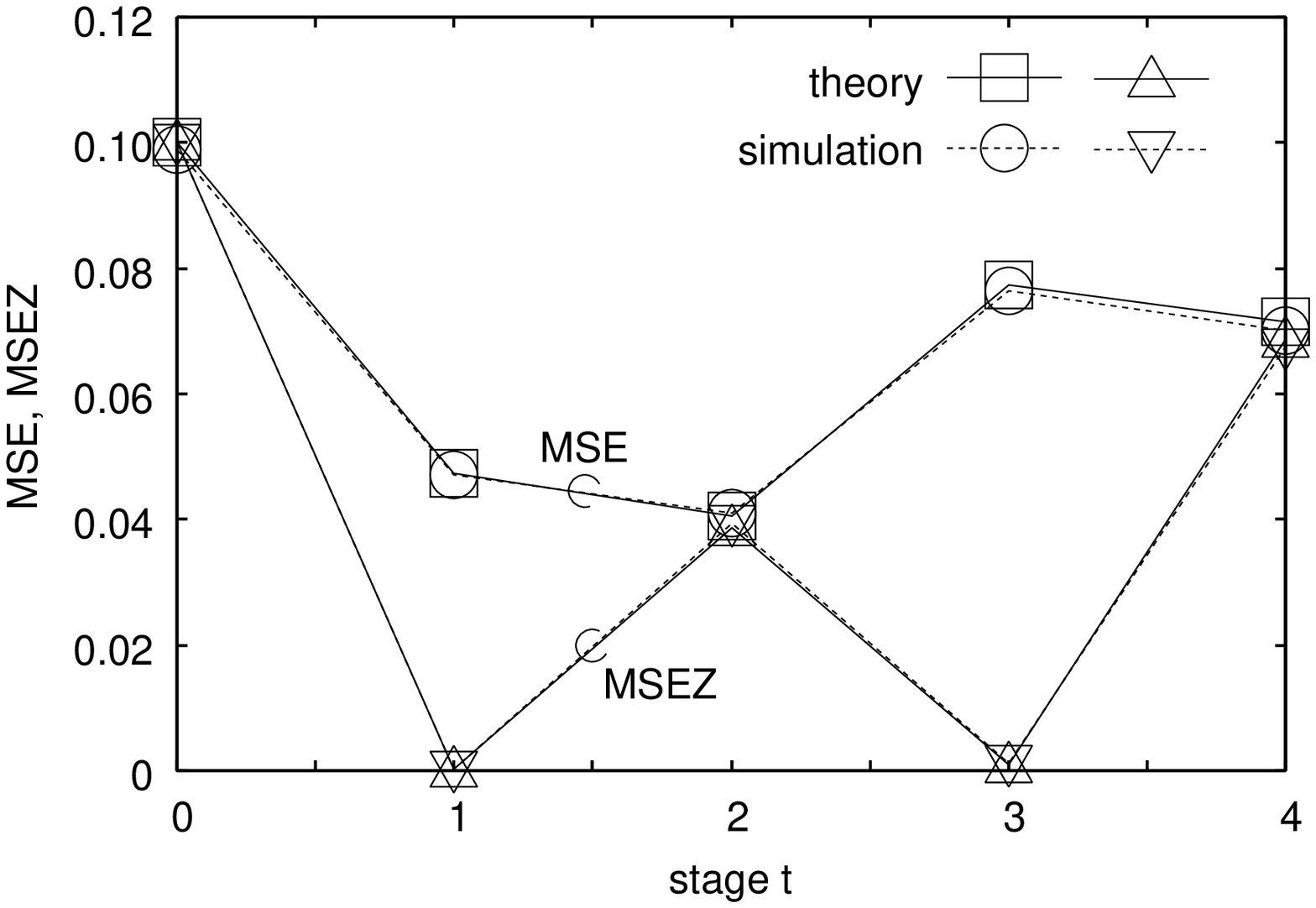} \\
    \footnotesize $\qquad$ (b) \normalsize \\[2mm]
    \includegraphics[width=.7\linewidth,keepaspectratio]{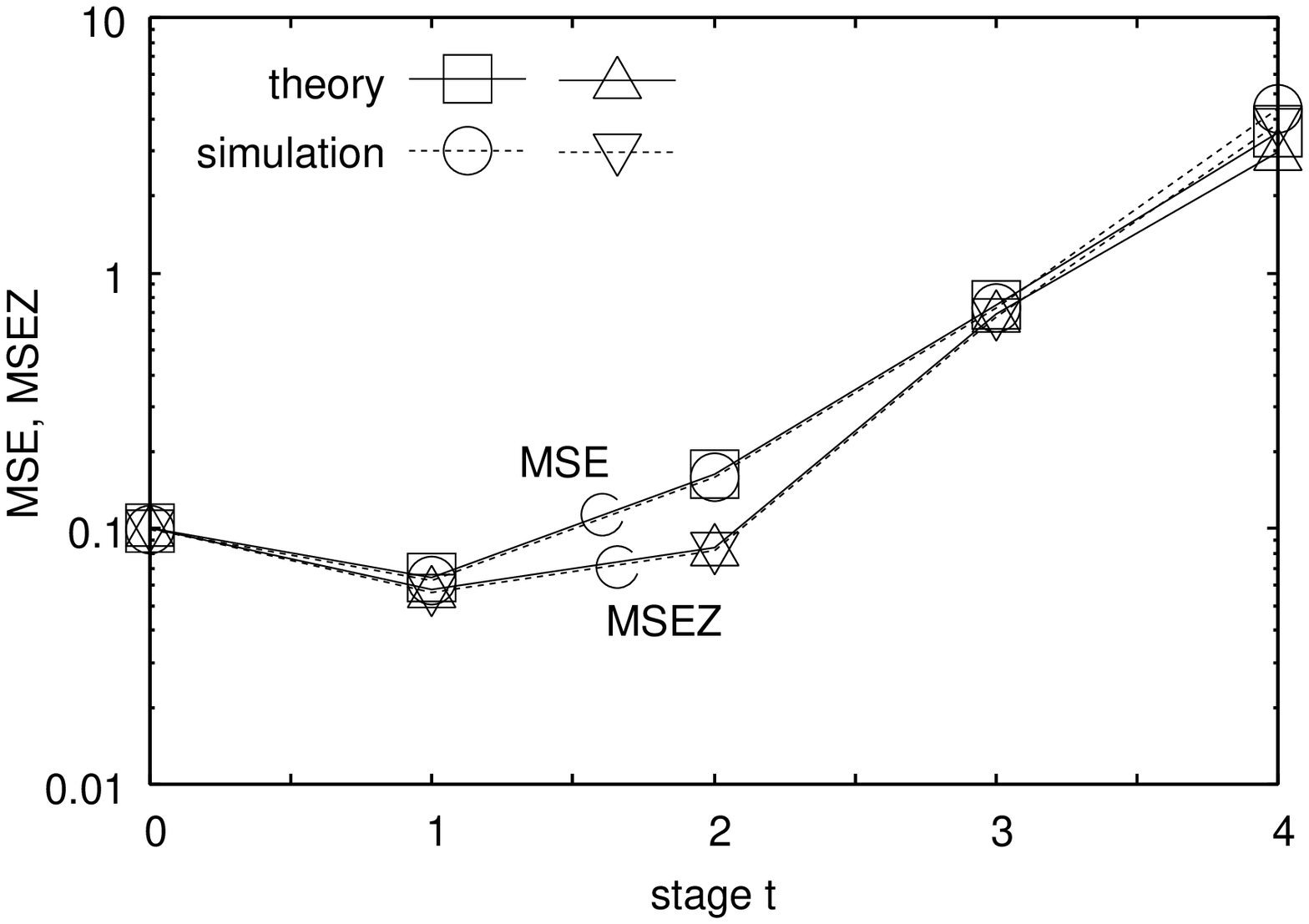} \\
    \footnotesize $\qquad$ (c) \normalsize \\[2mm]
    \caption{
      The first few stages of the dynamics of IST predicted by GFA (squares: MSE, trianlges: MSEZ). 
      Computer simulations (circles: MSE, Inverted triangles: MSEZ) are evaluated in $N=2,000$ systems. 
      (a) Case where the reconstruction succeeds. $(\rho, \delta, \lambda, c) = (0.1, 0.5, 3, 3)$. 
      (b) Failure case. $(\rho, \delta, \lambda, c) = (0.1, 0.8, 3, 1)$. 
      (c) Another failure case. $(\rho, \delta, \lambda, c) = (0.1, 0.8, 0.5, 1)$. 
    }
    \label{fig:result}
  \end{center}
\end{figure}

Figure \ref{fig:result} shows the first few stages of the dynamics of IST which is predicted by GFA. 
The parameters are set to be $(\rho, \delta, \lambda, c)$ $\in$ $\{(0.1, 0.5, 3, 3),$ $(0.1, 0.8, 3, 1),$ $(0.1, 0.8, 0.5, 1) \}$. 
The parameter $\lambda$ is not optimized for IST. 
Figure \ref{fig:result}(a) is a case where the reconstruction is successful. 
The parameter $c$ is set to be $c>1$ like the SSF method in this case. 
Since the residue is added little by little, it is easy to avoid a vibration behaviour. 
Figure \ref{fig:result}(b) is a case where the reconstruction fails. 
When the parameters are near the region where the reconstruction succeeds, 
a vibration behaviour often turns up. 
Figure \ref{fig:result}(c) is also a case where the reconstruction fails. 
When the parameters are far from the region where the reconstruction succeeds, MSE generally diverges. 
The GFA prediction is in good agreement with computer simulation result. 
The parameter $c$ is set to one in Fig. \ref{fig:result}(b) and Fig. \ref{fig:result}(c), 
which corresponds to the simple iterative thresholding algorithm (ITA) \cite{Donoho2009}. 
\par
The self-consistent equations appeared in Proposition {proposition:IST} 
involve three kinds of parameters $\{m^{(s)},$ $C^{(s,s')},$ $G^{(s,s')}\}_{s,s'=0}^{t}$. 
The number of these parameters is $t + 2 t^2$ and gradually grows as time passes. 
The other parameters can be easily calculated from these. 
When one solve self-consistent equations according to its definition, 
the computational cost for stage $t$ becomes $O(t^2 e^t)$ since each parameter, e.g., $m^{(t)}$, involves a $t$-multiple integral, 
Approximation schemes to evaluate the GFA result might be therefore important to capture long time dynamics \cite{Eissfeller1992, Heimel2001}. 

\section{Summary}
\par
We analyzed dynamics of the iterative shrinkage-thresholding algorithm for compressed sensing as a typical algorithm 
which cannot cancel the correlation between the present messages and their past values. 
While the state evolution plays an important role to understand nature of iterative algorithms 
which can cancel such a correlation exactly, 
the generating functional formalism gives us a analytical method to treat iterative algorithms which cannot cancel the correlation. 
The result of the generating functional formalism for algorithms which can cancel the correlation must give that of SE. 
It is under way to check this property.

\section*{Acknowledgment}
The author would like to thank Andrea Montanari for his valuable comments. 
This work was partially supported by 
a Grant-in-Aid for Scientific Research (C) No. 22500136 
from the Ministry of Education, Culture, Sports, Science and Technology (MEXT) of Japan.

\appendix

\subsection{Outline of analysis}
\par
Let $\vec{u}^{(t)}=(u_n^{(t)})$ be a summation of messages, i.e., 
$\vec{u}^{(t)} \triangleq \frac 1c A^\top \vec{z}^{(t)} + \vec{x}^{(t)} + \vec{\theta}^{(t)}$, 
where $\vec{\theta}^{(t)}$ is an external message which is introduced to evaluate the response function $G^{(s,s')}$. 
The Dirac's delta function is replaced as $\delta(x) = \gamma (2\pi)^{-1/2} e^{-\gamma^2 x^2/2}$ and 
the parameter $\gamma$ is taken the limit $\gamma \to \infty$ later. 
We first separate the summation of messages at any iteration step by inserting the following delta-distributions: 
$ 1 = \int \delta\vec{u}\delta\hat{\vec{u}} \prod_{s=0}^{t-1} \prod_{n=1}^N \exp [ i\hat{u}_n^{(s)} \{ u_n^{(s)} - ( \frac 1c A^\top \vec{z}^{(t)} )_n - x_n^{(t)} - \theta_n^{(t)} \} ]$, 
where $\delta\vec{u} \triangleq \prod_{s=0}^{t-1} \prod_{n=1}^N \frac{du_n^{(s)}}{\sqrt{2\pi}}$ 
and $\delta\hat{\vec{u}} \triangleq \prod_{s=0}^{t-1} \prod_{n=1}^N \frac{d\hat{u}_n^{(s)}}{\sqrt{2\pi}}$. 
Here, $( \vec{a} )_n$ denotes the $n^{\mathrm{th}}$ element of the vector $\vec{a}$. 
We then have 
\begin{align}
  \bar{Z}[\vec{\psi}]
  =& \mathbb{E}_{ \vec{x}_0, A, \vec{\omega} } \biggl\{ \sum_{\vec{x}^{(0)},\cdots,\vec{x}^{(t)}} \!\!\!\! p[\vec{x}^{(0)}] 
     \int_{\mathbb{R}^{2tN}} \!\!\!\! \delta\vec{u} \delta\hat{\vec{u}} 
     e^{- \rmi \sum_{s=0}^t \vec{x}^{(s)} \cdot \vec{\psi}^{(s)} } \nonumber \\
   & \times \exp \biggl[ \rmi \sum_{s=0}^{t-1} \sum_{n=1}^N 
     \hat{u}_n^{(s)} \{ u_n^{(s)} - x_n^{(s)} - \theta_n^{(s)} \} \nonumber \\
   & + \sum_{s=0}^t \sum_{n=1}^N \{ \ln \frac{\gamma}{\sqrt{2\pi}} 
     - \frac{\gamma^2}2 [ x_n^{(t+1)}-\eta_s(u_n^{(s)}) ]^2 \} \biggr] \nonumber \\
   & \times \exp \biggl[ - \rmi \frac 1c \sum_{m=1}^M \sum_{s=0}^{t-1} 
     \biggl( \sum_{n=1}^N a_{mn} \hat{u}_n^{(s)} \biggr) \omega_m \nonumber \\
   & - \rmi \frac 1c \sum_{m=1}^M \sum_{s=0}^{t-1} 
     \biggl( \sum_{n=1}^N a_{mn} \hat{u}_n^{(s)} \biggr) \nonumber \\
   & \times \biggl( \sum_{n=1}^N a_{mn} \{ x_{0,n} - x_n^{(s)} \} \biggr) \biggr] \biggr\} , 
\end{align}
In order to average the generating functional with respect to the disorder $A$ and $\vec{\omega}$, 
we isolate the spreading codes by introducing the variables $v_m^{(s)}, w_m^{(s)}$: 
$1 =$ $\int \delta\vec{v}\delta\hat{\vec{v}}$ $\prod_{s=0}^{t-1}$ $\prod_{m=1}^M$ 
      $\exp [ i\hat{v}_m^{(s)} \{ v_m^{(s)}$ $- \sqrt{\delta} \sum_{n=1}^N$ 
      $a_{mn} \hat{u}_n^{(s)} \} ]$ and 
$1 =$ $\int \delta\vec{w}\delta\hat{\vec{w}}$ $\prod_{s=0}^{t-1}$ $\prod_{m=1}^M$ 
      $\exp [ i\hat{w}_m^{(s)} \{ w_m^{(s)}$ $- \sqrt{\delta} \sum_{n=1}^N$ 
      $a_{mn} [ x_{0,n} -x_n^{(s)} ] \} ]$,
where
$\delta\vec{v} \triangleq \prod_{m=1}^{M} \prod_{s=0}^{t-1} \frac{dv_m^{(s)}}{\sqrt{2\pi}}$, 
$\delta\hat{\vec{v}} \triangleq \prod_{m=1}^{M} \prod_{s=0}^{t-1} \frac{d\hat{v}_m^{(s)}}{\sqrt{2\pi}}$, 
$\delta\vec{w} \triangleq \prod_{m=1}^{M} \prod_{s=0}^{t-1} \frac{dw_m^{(s)}}{\sqrt{2\pi}}$, and 
$\delta\hat{\vec{w}} \triangleq \prod_{m=1}^{M} \prod_{s=0}^{t-1} \frac{d\hat{w}_m^{(s)}}{\sqrt{2\pi}}$. 
One then obtains 
\begin{align}
   & \mathbb{E}_{A, \vec{\omega}} \biggl\{ \exp \biggl[ - \rmi \frac 1c \sum_{m=1}^M \sum_{s=0}^{t-1} 
     \biggl( \sum_{n=1}^N a_{mn} \hat{u}_n^{(s)} \biggr) \omega_m \nonumber \\
   & - \rmi \frac 1c \sum_{m=1}^M \sum_{s=0}^{t-1} 
     \biggl( \sum_{n=1}^N a_{mn} \hat{u}_n^{(s)} \biggr) \biggl( \sum_{n=1}^N a_{mn} \{ x_{0,n} - x_n^{(s)} \} \biggr) \biggr] \biggr\} \nonumber \\
  =& \int_{\mathbb{R}^{4tN}} \delta\vec{v}\delta\hat{\vec{v}}\delta\vec{w}\delta\hat{\vec{w}} \nonumber \\
   & \times \exp \biggl[ \rmi \sum_{m=1}^M \sum_{s=0}^{t-1} 
    \{ \hat{v}_m^{(s)} v_m^{(s)} + \hat{w}_m^{(s)} w_m^{(s)} - \frac1{c \delta} v_m^{(s)} w_m^{(s)} \} \biggr] \nonumber \\
   & \times \mathbb{E}_{ \vec{\omega} } \biggl\{ \exp \biggl[ 
     - \rmi \frac 1c \sqrt{\frac1\delta} \sum_{m=1}^M \sum_{s=0}^{t-1} v_m^{(s)} \omega_m \biggr] \biggr\} \nonumber \\
   & \times \mathbb{E}_{ A } \biggl\{ \exp \biggl[ 
     - \rmi \sqrt{\delta} \sum_{m=1}^M \sum_{s=0}^{t-1} 
     \{ \hat{v}_m^{(s)} \sum_{n=1}^N a_{mn} \hat{u}_n^{(s)} \nonumber \\
   & + \hat{w}_m^{(s)} \sum_{n=1}^N a_{mn} (x_{0,n}-x_n^{(s)}) \} \biggr] \biggr\} . \label{eq:average}
\end{align}
We now can calculate the average of the term in the disorder-averaged generating functional. 
The term $\mathbb{E}_{ \vec{\omega} } \{\cdots\}$ in (\ref{eq:average}) becomes 
\begin{align}
  & \mathbb{E}_{ \vec{\omega} } \biggl( \exp \biggl[ - \rmi \frac 1c \sqrt{\frac1\delta} \sum_{m=1}^M \sum_{s=0}^{t-1} v_m^{(s)} \omega_m \biggr] \biggr) \nonumber \\
  & = \exp \biggl[ -\frac{\sigma_{\omega}^2}{2 c^2 \delta} \sum_{m=1}^M \sum_{s=0}^{t-1} \sum_{s'=0}^{t-1} v_\mu^{(s)} v_\mu^{(s')} \biggr]. 
\end{align}
Calculating the average of the term containing the disorder in $\bar{Z}[\vec{\psi}]$ , 
we separate the relevant one-stage and two-stage order parameters by inserting: 
$1 =$ $(\frac N{2\pi})^{t}$ $\int d\vec{\mathsf{m}}d\hat{\vec{\mathsf{m}}}$ $\exp [ i N \sum_{s=0}^{t-1}$ $\hat{\mathsf{m}}^{(s)} \{ \mathsf{m}^{(s)}$ $- \frac1N$ $\sum_{n=1}^N$ $x_{0,n} x_n^{(s)} \} ]$, 
$1 =$ $(\frac N{2\pi})^{t}$ $\int d\vec{k}d\hat{\vec{k}}$ $\exp [ i N \sum_{s=0}^{t-1}$ $\hat{k}^{(s)} \{ k^{(s)}$ $- \frac1N$ $\sum_{n=1}^N$ $x_{0,n} \hat{u}_n^{(s)} \} ]$, 
$1 =$ $(\frac N{2\pi})^{t^2}$ $\int d\vec{q}d\hat{\vec{q}}$ $\exp [ i N \sum_{s=0}^{t-1} \sum_{s'=0}^{t-1}$ $\hat{q}^{(s,s')} \{ q^{(s,s')}$ $- \frac1N$ $\sum_{n=1}^N$ $x_n^{(s)} x_n^{(s')} \} ]$, 
$1 =$ $(\frac N{2\pi})^{t^2}$ $\int d\vec{Q}d\hat{\vec{Q}}$ $\exp [ i N \sum_{s=0}^{t-1} \sum_{s'=0}^{t-1}$ $\hat{Q}^{(s,s')} \{ Q^{(s,s')}$ $- \frac1N$ $\sum_{n=1}^N$ $\hat{u}_n^{(s)} \hat{u}_n^{(s')} \} ]$ and 
$1 =$ $(\frac N{2\pi})^{t^2}$ $\int d\vec{L}d\hat{\vec{L}}$ $\exp [ i N \sum_{s=0}^{t-1} \sum_{s'=0}^{t-1}$ $\hat{L}^{(s,s')} \{ L^{(s,s')}$ $- \frac1N$ $\sum_{n=1}^N$ $x_n^{(s)} \hat{u}_n^{(s')} \} ]$. 
Since the initial state probability is factorizable, 
the disorder-averaged generating functional factorizes into single-site contributions. 
\par
The disorder-averaged generating functional is for $N\to\infty$ dominated by a saddle-point \cite{Copson1965, Merhav2009}. 
We can thus simplify the saddle-point problem to (\ref{eq:Z3}). 
The disorder-averaged generating functional is then simplified to the saddle-point problem as 
\begin{align}
  \bar{Z}[\vec{\psi}] 
  =& \mathbb{E}_{\vec{x}_0} \biggl( \int 
  \rmd\vec{\mathsf{m}} d\hat{\vec{\mathsf{m}}}
  \rmd\vec{k} d\hat{\vec{k}}
  \rmd\vec{q} d\hat{\vec{q}}
  \rmd\vec{Q} d\hat{\vec{Q}}
  \rmd\vec{L} d\hat{\vec{L}} \nonumber \\
  & \times \exp \biggl[ N(\Psi+\Phi+\Omega)+O(\ln N) \biggr] \biggr), 
  \label{eq:Z3}
\end{align}
in which the functions $\Psi$, $\Phi$, $\Omega$ are given by 
\begin{align}
  \Psi \triangleq
  & i \sum_{s=0}^{t-1} \{ \hat{\mathsf{m}}^{(s)} \mathsf{m}^{(s)} + \hat{k}^{(s)}k^{(s)} \} + i \sum_{s=0}^{t-1} \sum_{s'=0}^{t-1}  \{ \hat{q}^{(s,s')} q^{(s,s')} \nonumber \\
  & + \hat{Q}^{(s,s')} Q^{(s,s')} + \hat{L}^{(s,s')} L^{(s,s')} \} 
\end{align}
\begin{align}
  \Phi \triangleq
  & \frac 1N \sum_{n=1}^N \ln \biggl\{ \int_{\mathbb{R}^{t+1}} \biggl( \prod_{s=0}^{t-1} x^{(s)} \biggr) p[x^{(0)}] \int \delta u \delta\hat{u} \nonumber \\
  & \times \exp \biggl[ \sum_{s=0}^{t-1} \{ \ln \frac{\gamma}{\sqrt{2\pi}} - \frac{\gamma^2}2 [x^{(s+1)}-\eta_s(u^{(s)})]^2 \} \nonumber \\
  & - i \sum_{s=0}^{t-1} \sum_{s'=0}^{t-1} \{ \hat{q}^{(s,s')} x^{(s)} x^{(s')} \nonumber \\
  & \quad + \hat{Q}^{(s,s')} \hat{u}^{(s)} \hat{u}^{(s')} + \hat{L}^{(s,s')} x^{(s)} \hat{u}^{(s')} \} \nonumber \\
  & + i \sum_{s=0}^{t-1} \hat{u}^{(s)} \{ u^{(s)} - x^{(s)} - \theta_n^{(s)} - x_{0,n} \hat{k}^{(s)} \} \nonumber \\
  & - i \sum_{s=0}^{t-1} x_{0,n} x^{(s)} \hat{\mathsf{m}}^{(s)} - i \sum_{s=0}^t x^{(s)} \psi_n^{(s)} \biggr] 
\end{align}
\begin{align}
  \Omega \triangleq
  & \frac 1N \ln \int \delta\vec{v}\delta\hat{\vec{v}}\delta\vec{w}\delta\hat{\vec{w}} \nonumber \\
  & \times \exp \biggl[ i \sum_{m=1}^M \sum_{s=0}^{t-1} \{ \hat{v}_m^{(s)} v_m^{(s)} + \hat{w}_m^{(s)} w_m^{(s)} - \frac1{c \delta} v_m^{(s)} w_m^{(s)} \} \nonumber \\
  & - \frac 12 \sum_{m=1}^M \sum_{s=0}^{t-1} \sum_{s'=0}^{t-1} \{ \frac1{c^2 \delta} \sigma_\omega^2 v_m^{(s)} v_m^{(s')} + \hat{v}_m^{(s)} Q^{(s,s')} \hat{v}_m^{(s')} \} \nonumber \\
  & - \frac 12 \sum_{m=1}^M \sum_{s=0}^{t-1} \sum_{s'=0}^{t-1} \{ \hat{v}_m^{(s)} [k^{(s)} - L(s',s)] \hat{w}_m^{(s')} \nonumber \\
  & \quad + \hat{w}_m^{(s)} [k^{(s')} - L^{(s,s')}] \hat{v}_m^{(s')} \} \nonumber \\
  & - \frac 12 \sum_{m=1}^M \sum_{s=0}^{t-1} \sum_{s'=0}^{t-1} \! \{ \! \hat{w}_m^{(s)} [ \rho - \! \mathsf{m}^{(s)} \! - \! \mathsf{m}^{(s')} \! + \! q^{(s,s')}] 
  \hat{w}_m^{(s')} \} \biggr]
\end{align}
where $\delta u \triangleq \prod_{s=0}^{t-1} \frac{du^{(s)}}{\sqrt{2\pi}}$ and $\delta \hat{u} \triangleq \prod_{s=0}^{t-1} \frac{d\hat{u}^{(s)}}{\sqrt{2\pi}}$. 
In the limit $N\to\infty$, the integral (\ref{eq:Z3}) will be dominated by the saddle point of the extensive exponent $\Psi+\Phi+\Omega$. 
\par
One can deduce the meaning of order parameter 
by derivation of the averaged generating functional $\bar{Z}[\vec{\psi}]$ 
with respect to the external messages $\{\theta_n^{(s)}\}$ and the dummy functions $\{\psi_n^{(s)}\}$. 
The averaged generating functional $\bar{Z}[\vec{\psi}]$ is dominated by a saddle-point for $N\to\infty$. 
We can thus simplify (\ref{eq:Z3}) in the large system limit. 
Using $\bar{Z}[\vec{0}]=1$, 
From derivatives of the averaged generating functional, we find
\begin{align}
  \mathbb{E}_{\vec{x},A,\vec{\omega}}(x_n^{(s)})
  =& \langle x^{(s)} \rangle_n, \label{eq:fd1} \\
  \mathbb{E}_{\vec{x},A,\vec{\omega}} (x_n^{(s)} x_{n'}^{(s')})
  =& \delta_{n,n'} \mathbb{E}_{\vec{x}_0} \langle x^{(s)}x^{(s')}\rangle_n, \nonumber \\
   & + (1-\delta_{n,n'}) \langle x^{(s)}\rangle_n \mathbb{E}_{\vec{x}_0} \langle x^{(s')}\rangle_{n'} \\
  \frac{\partial \mathbb{E}_{\vec{x},A,\vec{\omega}}(x_n^{(s)})}{\partial \theta_{n'}^{(s')}}
  =& - \rmi \delta_{n,n'} \langle x^{(s)}\hat{u}^{(s')}\rangle_n, \label{eq:fd3}
\end{align}
where $\langle \cdots \rangle_n$ denotes the average as 
$\langle f(\vec{x},\vec{u},\hat{\vec{u}}) \rangle_n$ $\triangleq$ 
$[\sum_{x^{(0)},\cdots,x^{(t)}}$ $\int \delta u\delta \hat{u}$ $\mu_n(\vec{x},\vec{u},\hat{\vec{u}})$ $f(\vec{x},\vec{u},\hat{\vec{u}})]/$
$[\sum_{x^{(0)},\cdots,x^{(t)}}$ $\int \delta u\delta \hat{u}$ $\mu_n(\vec{x},\vec{u},\hat{\vec{u}})]$ 
with 
$\mu_n(\vec{x},\vec{u},\hat{\vec{u}})$ $\triangleq$ 
$\delta [x^{(0)}]$ $\exp [$
$\sum_{s=0}^{t-1}$ $\{ \ln \frac{\gamma}{\sqrt{2\pi}}$ $- \frac{\gamma^2}2 [x^{(s+1)}-\eta_s(u^{(s)})]^2 \}$ 
$- \rmi \sum_{s=0}^{t-1}$ $\sum_{s'=0}^{t-1}$ $\{ \hat{q}^{(s,s')}$ $x^{(s)}$ $x^{(s')}$ 
$+ \hat{Q}^{(s,s')}$ $\hat{u}^{(s)}$ $\hat{u}^{(s')}$ 
$+ \hat{L}^{(s,s')}$ $x^{(s)}$ $\hat{u}^{(s')} \}$
$+ \rmi \sum_{s=0}^{t-1}$ $\hat{u}^{(s)}$ $\{ u^{(s)}$ $- x^{(s)}$ $- \theta_n^{(s)}$ $- x_{0,n} \hat{k}^{(s)} \}$ 
$- \rmi \sum_{s=0}^{t-1}$ $x_{0,n} x^{(s)}$ $\hat{\mathsf{m}}^{(s)} ]$ $|_{\mathrm{saddle}}$. 
Here, $f|_{\mathrm{saddle}}$ denotes an evaluation of a function $f$ at the dominating saddle-point. 
The saddle-point equations are derived by differentiation of $N(\Psi+\Phi+\Omega)$ 
with respect to integration variables 
$\{\vec{m},\hat{\vec{m}}$, 
$\vec{k},\hat{\vec{k}}$, 
$\vec{q},\hat{\vec{q}}$, 
$\vec{Q},\hat{\vec{Q}}$, 
$\vec{L}$ and $\hat{\vec{L}}\}$. 
These equations will involve the average overlap $m^{(s)}$, 
the average single-user correlation $C^{(s,s')}$ and the average single-user response function $G^{(s,s')}$: 
\begin{align}
  m^{(s)}    \triangleq & 
  \lim_{N \to \infty} \frac 1N \sum_{n=1}^N 
  \mathbb{E}_{\vec{x},A,\vec{\omega}}(x_{0,n} x_n^{(s)}), \\
  C^{(s,s')} \triangleq & 
  \lim_{N \to \infty} \frac 1N \sum_{n=1}^N 
  \mathbb{E}_{\vec{x},A,\vec{\omega}} (x_n^{(s)}x_{n'}^{(s')}), \\
  G^{(s,s')} \triangleq & 
  \lim_{N \to \infty} \frac 1N \sum_{n=1}^N 
  \frac {\partial \mathbb{E}_{\vec{x},A,\vec{\omega}} ( x_n^{(s)})}
        {\partial \theta_{n'}^{(s')}}. 
\end{align}
\par
Using the derivatives (\ref{eq:fd1}) -- (\ref{eq:fd3}), 
straightforward differentiation of $\Psi+\Phi+\Omega$ with respect to 
$m^{(s)}$, $\hat{\mathsf{m}}^{(s)}$, 
$k^{(s)}$, $\hat{k}^{(s)}$, 
$q^{(s,s')}$, $\hat{q}^{(s,s')}$, 
$Q^{(s,s')}$, $\hat{Q}^{(s,s')}$, 
$L^{(s,s')}$ and $\hat{L}^{(s,s')}$ 
leads us to the following saddle-point equations: 
$\hat{\mathsf{m}}^{(s)} =$ $\rmi \frac{\partial \Omega}{\partial m^{(s)}} |_{\mathrm{saddle}}$, 
$\hat{k}^{(s)} =$ $\rmi \frac{\partial \Omega}{\partial k^{(s)}} |_{\mathrm{saddle}}$, 
$m^{(s)} =$ $\lim_{N \to \infty}$ $\frac 1N$ $\sum_{n=1}^N$ $\langle x_{0,n} x^{(s)} \rangle_n$, 
$k^{(s)} =$ $0$, 
$\hat{q}^{(s,s')} =$ $\rmi \frac{\partial \Omega}{\partial q^{(s,s')}} |_{\mathrm{saddle}}$, 
$q^{(s,s')} =$ $\lim_{N \to \infty}$ $\frac 1N$ $\sum_{n=1}^N$ $\langle x^{(s)}x^{(s')} \rangle_n$, 
$\hat{Q}^{(s,s')} =$ $\rmi \frac{\partial \Omega}{\partial Q^{(s,s')}} |_{\mathrm{saddle}}$, 
$Q^{(s,s')} = 0$, 
$\hat{L}^{(s,s')} =$ $\rmi \frac{\partial \Omega}{\partial L^{(s,s')}} |_{\mathrm{saddle}}$ and 
$L^{(s,s')} =$ $\lim_{N \to \infty}$ $\frac 1N$ $\sum_{n=1}^N$ $\langle x^{(s)}\hat{u}^{(s')}\rangle_n$ 
for all $s$ and $s'$. 
We then find $\mathsf{m}^{(s)} = m^{(s)}$, $q^{(s,s')} = C^{(s,s')}$ and $L^{(s,s')} = \rmi G^{(s,s')}$. 
It should be noted that the causality $\partial \langle x^{(s)} \rangle / \partial \theta^{(s')}=0$, 
should be hold for $s \le s'$, therefore $L^{(s,s')}=G^{(s,s')}=0$ for $s \le s'$. 
\par
Straightforward differentiation and taking the limit $\gamma \to \infty$, 
we then arrive at Proposition \ref{proposition:IST}.

\end{document}